\documentclass[slac_one]{revtex4}
\usepackage{graphicx}
\usepackage{fancyhdr}
\usepackage{xspace}
\newcommand{\Dzero}{D\O \xspace}
\newcommand{\ttbar}{t\bar{t}}
\newcommand{\qqbar}{q\bar{q}}
\newcommand{\ppbar}{p\bar{p}}

\begin{document}

\title{Top Quark Properties from Top Pair Events and Decays} 

%

\author{A. Ivanov (for the CDF and \Dzero  Collaborations)}
\affiliation{University of California, Davis, CA, 95616, USA }

\begin{abstract}
Over a decade since the discovery of the top quark we are still trying to unravel
mysteries of the heaviest observed particle and learn more about its nature. 
The continuously accumulating statistics of CDF and \Dzero 
data provide the means for measuring 
top quark properties with ever greater precision 
and the opportunity to search for 
signs of new physics that could be manifested 
through subtle deviations 
from the standard model in the production and decays of top quarks.
In the following we present a slice of the rich program in top quark physics at the 
Fermilab Tevatron: measurements of the properties of top quark decays and
searches for unusual phenomena in events with pair produced tops.
In particular, we discuss the most recent and precise CDF and \Dzero 
measurements of the transverse polarization of W bosons from top decays, 
branching ratios and searches for flavor-changing neutral current decays, decays 
into charged Higgs and invisible decays. 
These analyses correspond to integrated luminosities
ranging from 0.9 to 2.7 fb$^{-1}$.
\end{abstract}

\maketitle

\thispagestyle{fancy}


\section{INTRODUCTION} 

The Tevatron $\ppbar$ collider operates 
at a center-of-mass energy $\sqrt{s} = 1.96$ TeV.
Top quarks are produced mainly strongly in pairs and,
within the standard model (SM), decay almost 
exclusively into a $W$ boson and a $b$-quark
via the weak interaction. Therefore, any deviation 
in the V-A structure of the $t \to Wb$ decay or
evidence for more exotic final states would be indicative
of new physics.  
Both the CDF and \Dzero Collaborations 
study the detailed kinematics of the products from $t \to Wb$ decay 
and search for non-SM contributions.
These will be summarized in this report.

\section{MEASUREMENT OF $W$ HELICITY IN TOP DECAYS}

The helicity of the $W$ bosons in top decays is defined by the V-A structure of the $tWb$ vertex 
and can be accessed through the kinematics of the $W$ decay products. 
The alignment of the $W$ spin
can be characterized by the angle~$\theta^\star$ 
between the $W$ momentum vector in the top quark rest frame and the 
momentum of the down-type fermion in the $W$ rest frame. 
The SM predicts the fraction of longitudinally polarized $W$ bosons
to be $f_0 \approx 70\%$, the fraction of left-handed $W$s $f_-\approx 30\%$, 
while the right-handed fraction $f_+$ is greatly suppressed by the V-A structure of the decay.

The CDF Collaboration has performed three independent measurements 
of the longitudinal $f_0$ and right-handed $f_+$ fractions of the helicity 
of the $W$ boson 
using  lepton $+$ jets events with at least one secondary-vertex tag ($b$-tag)
in 1.9~fb$^{-1}$ of data. 
The first method relies on the matrix element technique~\cite{me}. 
In this measurement, the fractions 
are set to $f_{+} = 0$   and  $f_- = 1-f_{0}$. 
The two other techniques focus on measuring the $\cos\theta^\star$ distribution
in kinematically reconstructed  $\ttbar$ final states.
In one case (``unfolding technique")
the observed $\cos\theta^\star$ distribution is corrected
by subtracting the background and accounting 
for bin-to-bin migration of $\ttbar$ events from true to reconstructed values of
$\cos\theta^\star$.  In the other analysis (``template method")
templates of $\cos{\theta^\star}$ 
for $\ttbar$ with different $W$ polarizations 
and backgrounds after event
reconstruction are fitted to the data.
Both analyses perform 1-parameter fits (fixing $f_0$ or $f_+$ to their SM values)
and 2-parameter fits by simultaneously extracting the $f_0$ and $f_+$ fractions.
The results of these two analyses have been further combined using the 
 BLUE technique~\cite{blue}.
The summary of CDF results is presented in Table~\ref{tab:whel}. 

The \Dzero Collaboration measures the $W$ polarization 
using the template technique in 2.7 fb$^{-1}$ of data
in both dilepton and lepton $+$ jets event topologies.
The lepton $+$ jets events are fully reconstructed,  and the information 
in the $W \to \qqbar$ decays is also sampled by randomly picking 
the down-type quark as one
of the jets associated with the $W$ boson.
Dileptons are reconstructed within their eight-fold ambiguity, and the average
of $\cos\theta^\star$
for all solutions is used in the extraction of helicity.
Results of  \Dzero measurement are also given in Table~\ref{tab:whel}. 
All results are consistent with their SM values.

\begin{table}[tb]
\begin{center}
\caption{Summary of $W$ helicity measurements.}
\begin{tabular}{lcccc}
\hline \textbf{Experiment}  & \textbf{Channel}  & \textbf{Technique} & \textbf{$f_0$} &  \textbf{$f_+$}  \\  \hline\hline
CDF & $\ell + $jets & Matrix Element  & 0.64  $\pm$ 0.08 (stat)  $\pm$ 0.07  (syst)   & fixed to 0.0 \\  \hline
CDF  & $\ell + $jets & $\cos{\theta^\star}$ Unfolding &  0.38  $\pm$ 0.21 (stat)  $\pm$ 0.07 (syst)   &  0.15  $\pm$ 0.10 (stat)   $\pm$ 0.04  (syst)    \\
& $\ell + $jets & fixed $f_0$ or $f_+$ & 0.66  $\pm$ 0.10 (stat) $\pm$ 0.06 (syst) & 0.01  $\pm$ 0.05  (stat) $\pm$ 0.03 (syst) \\   \hline

CDF  & $\ell + $jets & $\cos{\theta^\star}$ Template &  0.65  $\pm$ 0.19  (stat)  $\pm$ 0.03  (syst)   &  $-$0.03  $\pm$ 0.07 (stat)   $\pm$ 0.03  (syst)    \\ 
& $\ell + $jets & fixed $f_0$ or $f_+$ & 0.59  $\pm$ 0.11 (stat) $\pm$ 0.04 (syst) & $-$0.04  $\pm$ 0.04  (stat) $\pm$ 0.03 (syst) \\   \hline
	 CDF  & $\ell + $jets &  $\cos{\theta^\star}$ Combination &  0.66  $\pm$ 0.16   &  $-$0.03  $\pm$ 0.07    \\  
	 & $\ell + $jets & fixed $f_0$ or $f_+$ & 0.62  $\pm$ 0.11  & $-$0.04  $\pm$ 0.05   \\   \hline
	\Dzero   & $\ell\ell, \ell + $jets & $\cos{\theta^\star}$ Template &  0.49  $\pm$ 0.11  (stat)  $\pm$ 0.09 (syst)  &  0.11    $\pm$ 0.06  (stat)  $\pm$ 0.05  (syst)    \end{tabular}
\label{tab:whel}
\end{center}
\end{table}

\section{MEASUREMENT OF $R = {\cal B}(t \to Wb) / {\cal B} (t \to Wq )$ }

Within the SM, the top quark decays to a $W$ boson and a down-type quark $q$ $(q = d,s,b)$
with a rate proportional to $|V_{tq}|^2$. The ratio $R$ of the branching fractions 
of top decays to $Wb$ relative to $Wq$ can be expressed in terms of the 
CKM matrix elements as

\begin{equation}
\label{eq:R}
	R = \frac{{\cal B}(t \to W^+ b)}{{\cal B}(t \to W^+ q)} = \frac{|V_{tb}|^2}{|V_{td}|^2 + |V_{ts}|^2 + |V_{tb}|^2 }
\end{equation}

The average number of $b$-tagged jets in a $\ttbar$ event depends directly on the 
value of R and the probability of tagging a $b$-quark. The \Dzero 
collaboration measured $R$ simultaneously with the $\ttbar$ cross section ($\sigma_{\ttbar}$),
basing their result on the observed
 multiplicity distribution of $b$-tags shown in Fig.~\ref{fig:D0higgs}. A simultaneous likelihood fit
to these two variables yields $R = 0.97$ $^{+ 0.09}_{-0.08}$ and $\sigma_{\ttbar} = 8.18 ^{+0.90}_{-0.84}$ (stat + syst) $\pm$ 0.50 (lumi). The observed value of R is translated to a lower limit at 95\% confidence level using the Feldman-Cousins ordering principle~\cite{FC}, giving $R > 0.79$ at 95\% C.L. Assuming a unitary CKM matrix with three fermion generations yields $|V_{tb}| > 0.89$ at 95\% C.L.

\section{SEARCH FOR CHARGED HIGGS}

Charged Higgs $H^\pm$ bosons are predicted in supersymmetric and GUT extensions of the SM.
If a charged Higgs boson is sufficiently light, it can be produced in top quark decays.
In the presence of a charged Higgs boson, the $t \to H^+b$ decay would compete with the SM top 
quark decay $t \to W^+b$, thereby altering the expected number of events in different final states of $\ttbar$.
The event migration among different final states depends on the decays of $H^\pm$.
Within the MSSM, $H^+ \to c\bar{s}$ dominates at low $\tan \beta$, while $H^+ \to \tau\nu$
dominates at high $\tan \beta$.

\begin{figure}[t]
\includegraphics[width=5cm]{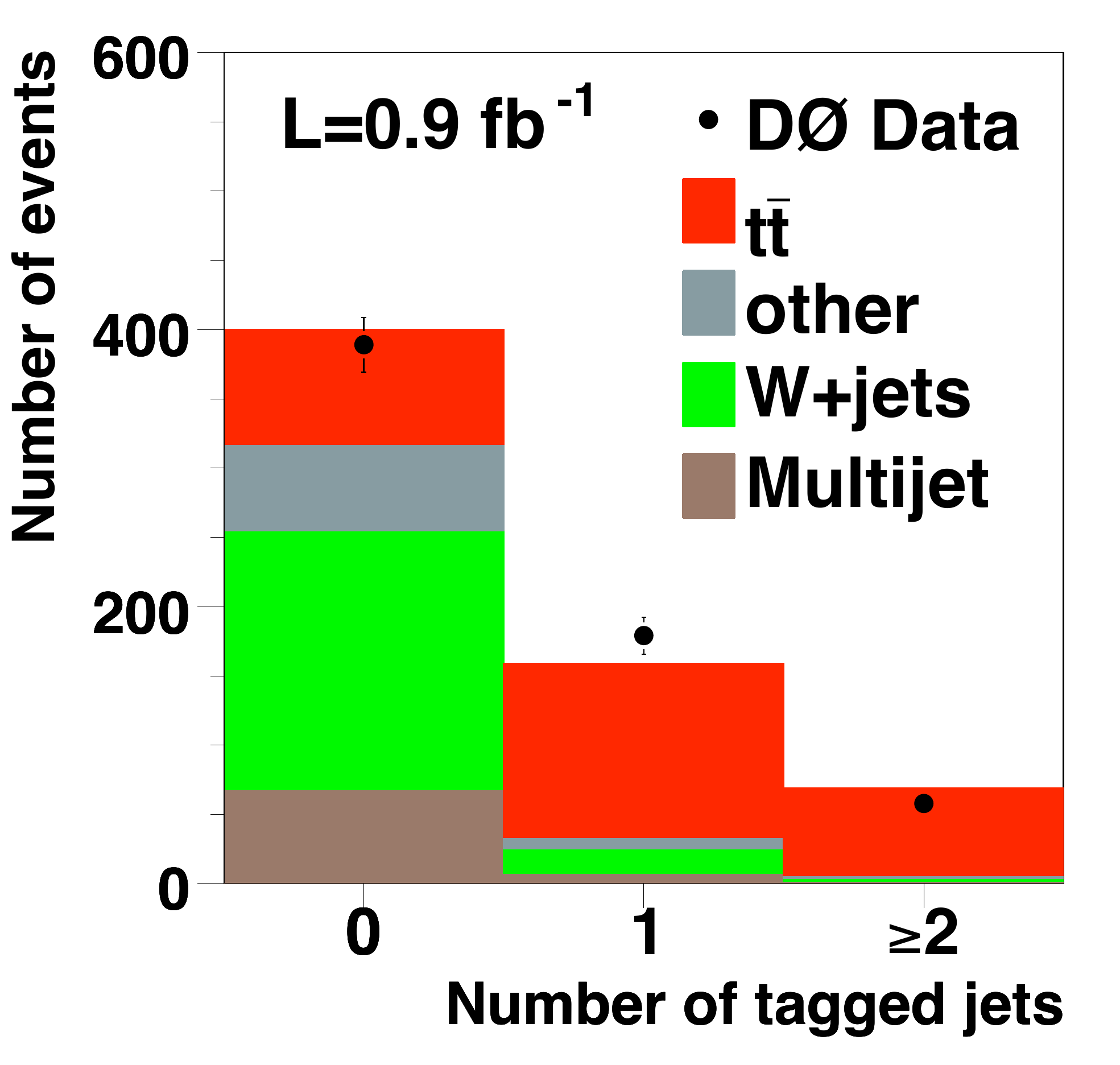} 
\includegraphics[width=5cm]{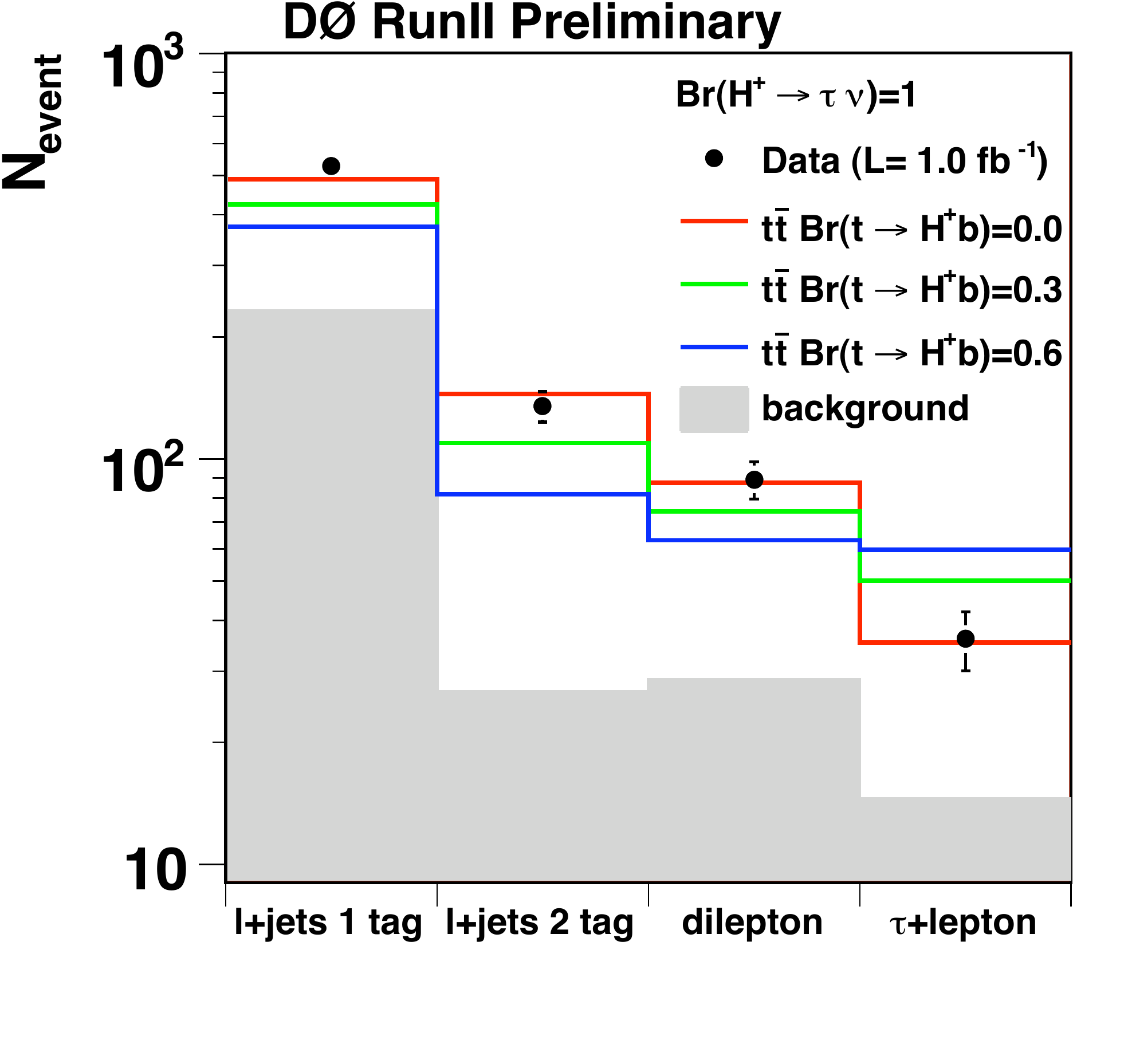}
\includegraphics[width=7cm]{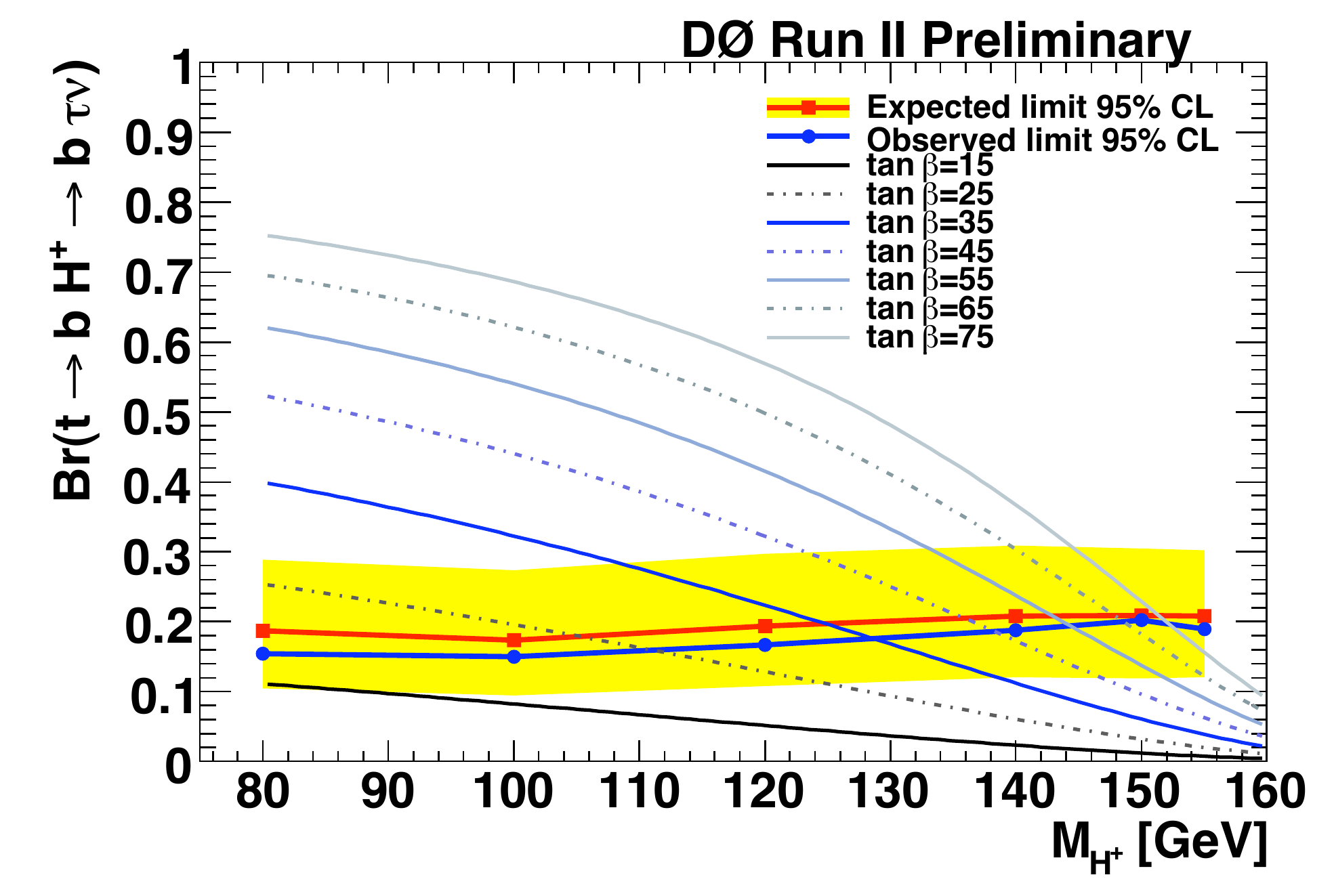}
\caption{\textbf{LEFT:} Predicted and observed 
number of events with 0, 1 and $\geq 2$ $b$-tags in the sample of lepton $+ \geq 4 $ jets. 
\textbf{CENTER:} Expected (for several ${\cal B}(t \to H^+b)$) and observed number 
of events in final states of $\ttbar$ decays.
\textbf{RIGHT:} The upper limit on ${\cal B}(t \to H^+ b \to \tau \nu b)$ at 95\% C.L. 
as a function of charged Higgs mass.
}
\label{fig:D0higgs}
 \end{figure}

\begin{figure}
\includegraphics[width=6.8cm]{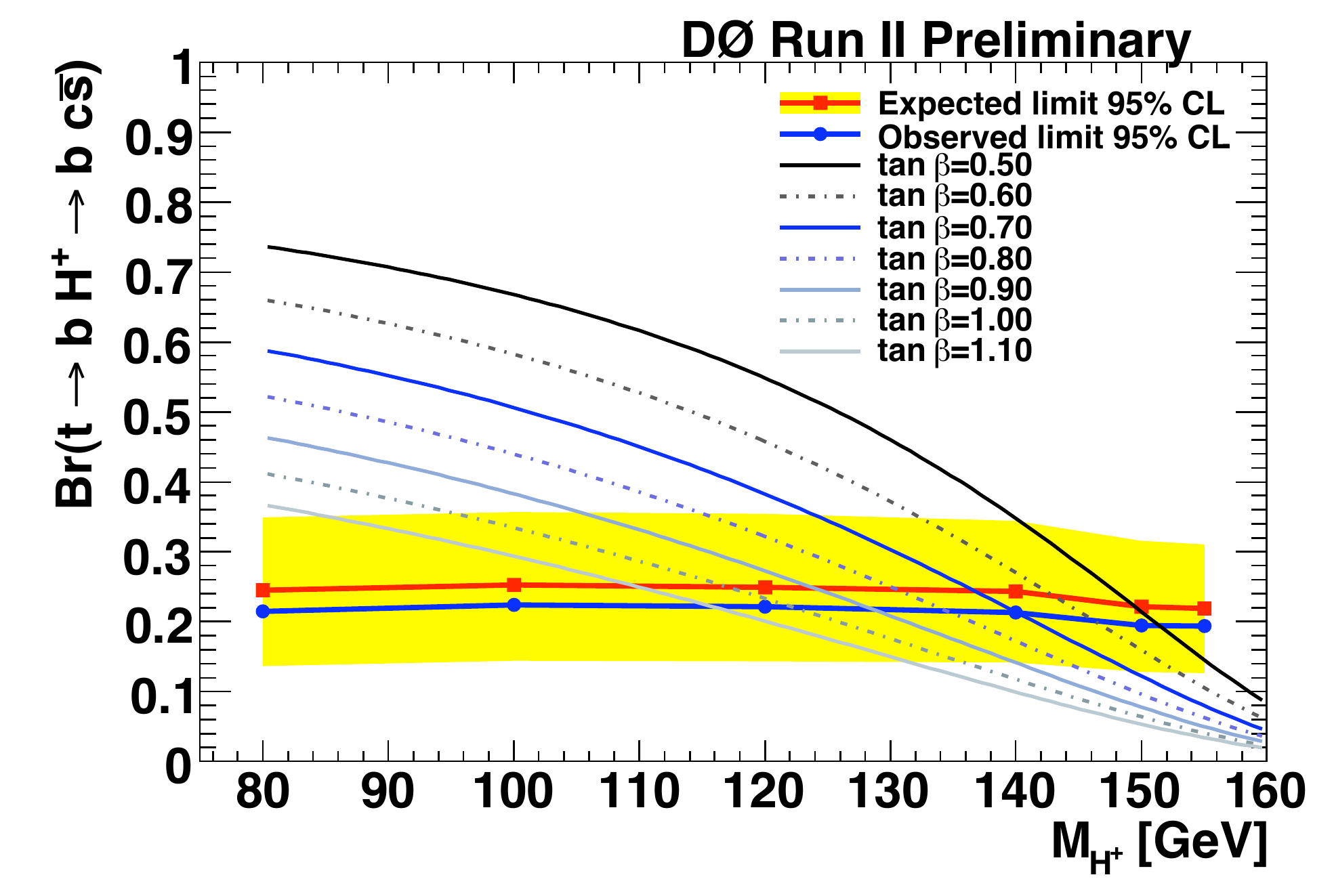}
\includegraphics[width=5.4cm]{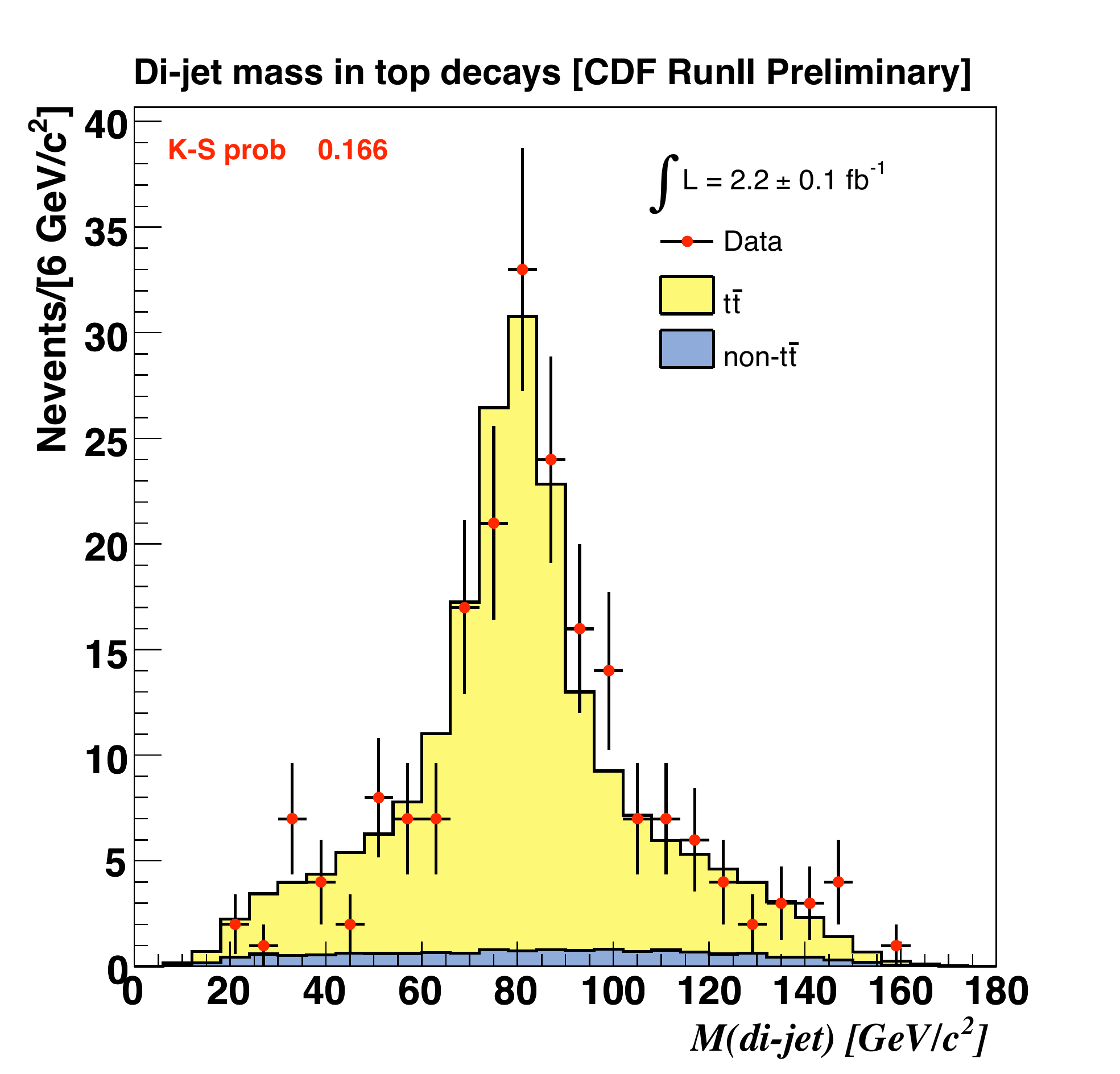}
\includegraphics[width=5.4cm]{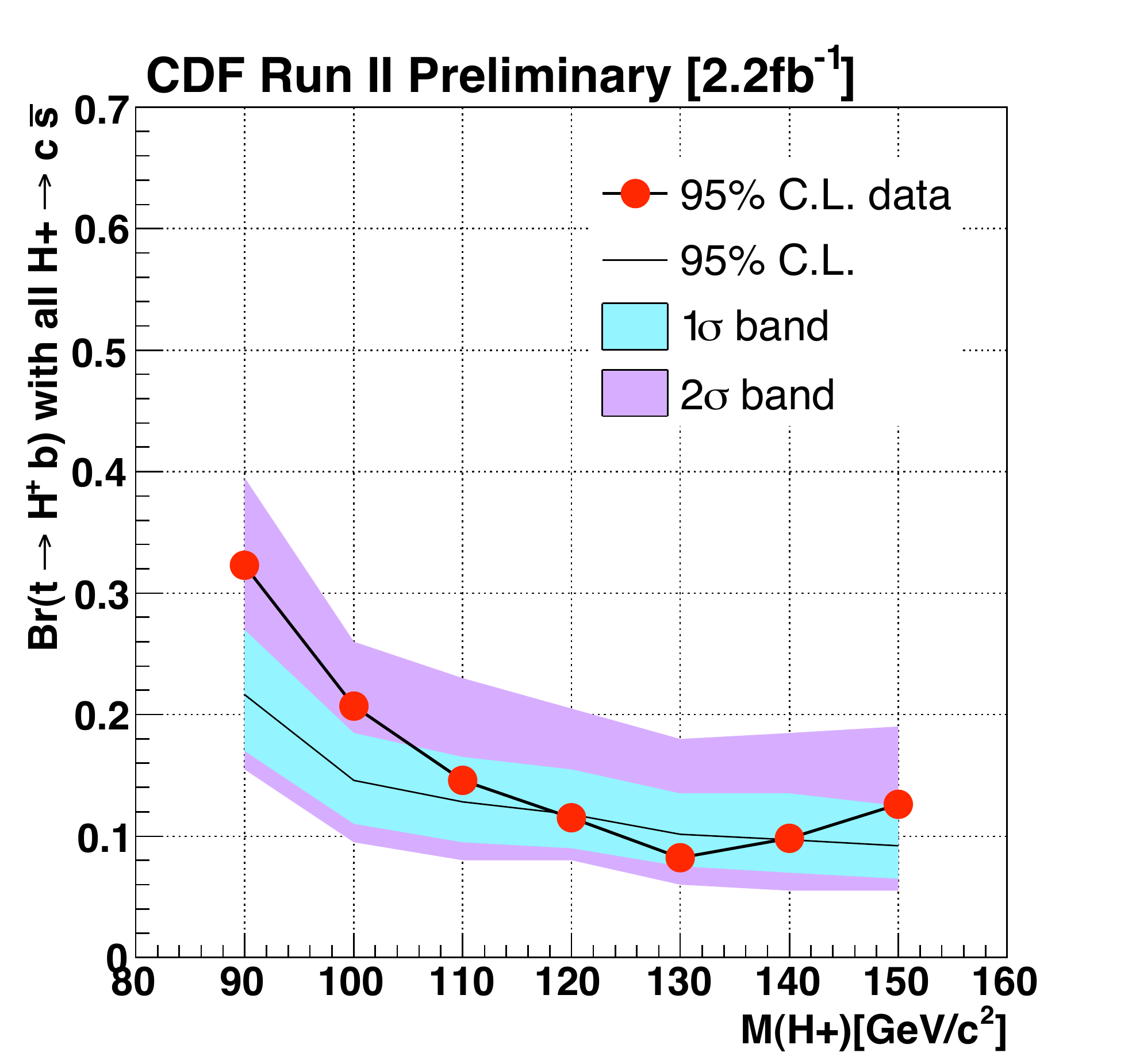}
\caption{
\textbf{LEFT:} The upper limit on the ${\cal B}(t \to H^+ b \to c \bar{s} b)$ at 95\% C.L. 
as a function of charged Higgs mass.
\textbf{CENTER:} Distribution of di-jet mass in $\ttbar$ decays. 
 \textbf{RIGHT:} The upper limit on the ${\cal B}(t \to H^+ b)$ at 95\% C.L.
 as a function of $m_{H^\pm}$. 
}
\label{fig:CDFhiggs}
 \end{figure}

The \Dzero Collaboration has considered the ${\cal B} (H^+ \to \tau\nu) = 100\%$ and 
``leptophobic" ${\cal B} (H^+ \to c\bar{s}) = 100\%$ scenarios, and measured the $\ttbar$ event yields 
in the dilepton ($\ell = e, \mu$), the $\tau +$ lepton, and the lepton $+$ jets channels, which are
constructed to be orthogonal to each other.
For different values of ${\cal B} (t \to H^+ b)$ the expected and observed numbers of events 
in the explored final states are shown in Fig.~\ref{fig:D0higgs}. 
No indication was found for charged Higgs boson production in the $\tau$ or leptophobic mode.
 The 95\% C.L. limits on ${\cal B} (t \to H^+ b \to \tau\nu b)$ and 
 ${\cal B} (t \to H^+ b \to c\bar{s} b)$ as a function of $H^\pm$ mass
  are presented in Fig.~\ref{fig:D0higgs} and
Fig.~\ref{fig:CDFhiggs} respectively.

The CDF Collaboration has also searched for the decays
$t \to H^+b \to  c\bar{s}b$ 
in the lepton $+$ jets events by fully reconstructing $\ttbar$ decay and 
exploiting the difference between 
the dijet mass spectra in $W \to q\bar{q^\prime}$ 
and $H^+ \to c\bar{s}$ decays. The invariant dijet mass spectrum in data
is presented in Fig.~\ref{fig:CDFhiggs}.
No significant deviation from the SM is observed, and limits are set 
on the ${\cal B} ( t\to H^+b \to c{\bar s} b$) as  a function of the charged Higgs mass shown in
Fig.~\ref{fig:CDFhiggs}.

\section{SEARCH FOR FLAVOR-CHANGING NEUTRAL CURRENT DECAYS $t \to Zq$ }

In the SM,  flavor-changing neutral current (FCNC) decays are highly suppressed, 
with expected branching fractions ${\cal B} ( t \to Zq)  \approx {\cal O} (10^{-14})$. 
 In SUSY and two-Higgs doublet models
branching ratios are higher, up to ${\cal O} (10^{-2})$~\cite{fcnc_theory}.

The CDF Collaboration has performed a search for FCNC top decays $t \to Zq$  using 
$Z \to \ell^+\ell^- (\ell= e, \mu)$ events with four and more jets and optimized selection. 
Events in the signal region are further classified according to availability of a secondary 
vertex ($b$-tag). A third sample is used as a control region and consists of rejected events 
that failed at least one of the optimized selection criteria. 

The signal is discriminated from background by exploring kinematic constraints present 
in FCNC events: $\ttbar \to WbZq \to \qqbar b \ell^+\ell^- q$. A $\chi^2$ variable is defined 
using reconstructed $b\qqbar$ and $q\ell^+\ell^-$ masses, their uncertainties and the accepted
values of the top and $W$ mass:

\begin{equation}
\label{eq:fcnc}
\chi^2 = \left( \frac{m_{\qqbar,rec} - m_W}{\sigma_{W \to \qqbar}}\right )^2 
+ \left( \frac{m_{b\qqbar,rec} - m_t}{\sigma_{t \to b\qqbar}}\right )^2 
+ \left( \frac{m_{q\ell^+\ell^-,rec} - m_t}{\sigma_{t \to q\ell^+\ell^-}}\right )^2 .
\end{equation}

\begin{figure}
\includegraphics[width=9cm]{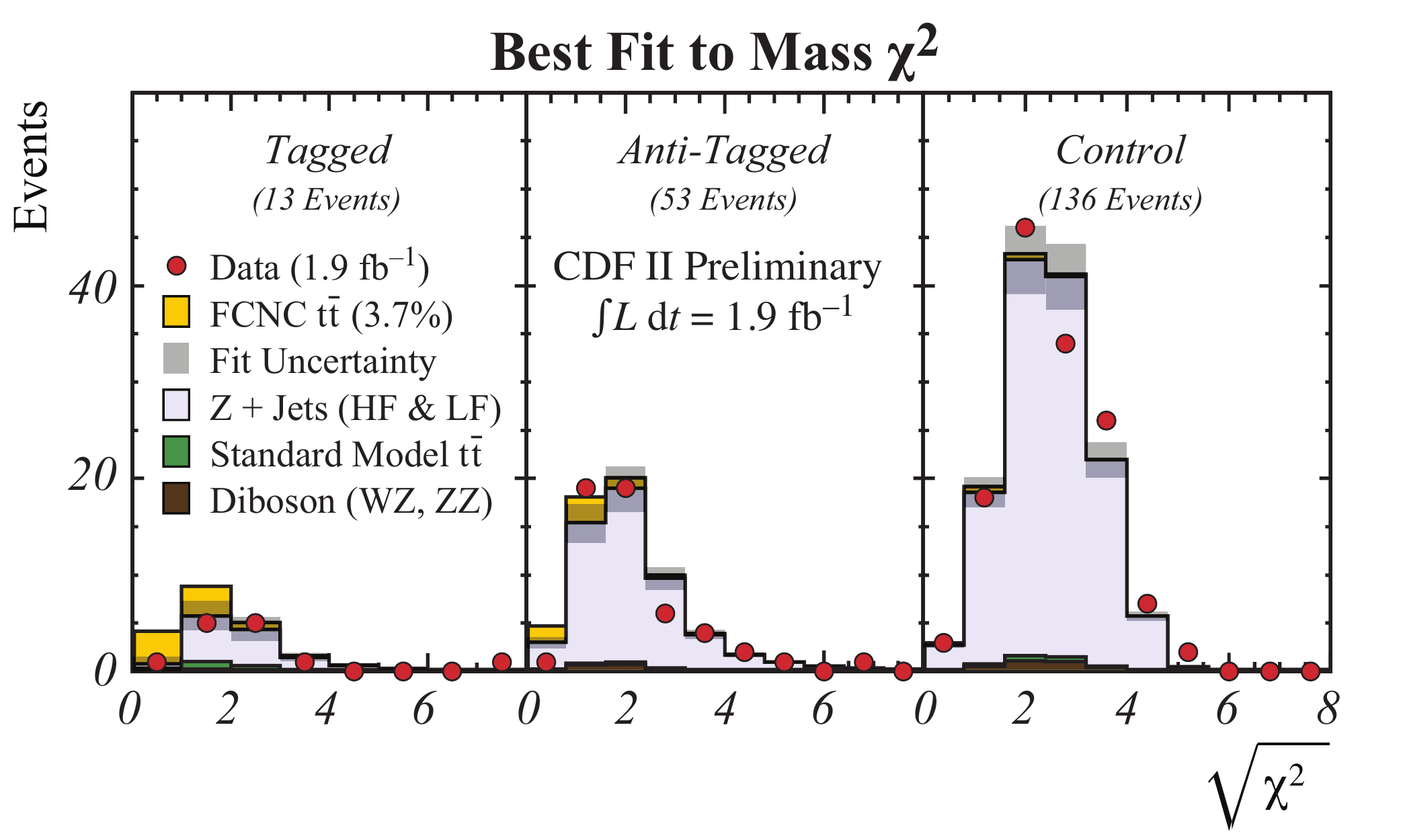}
\includegraphics[width=7.5cm]{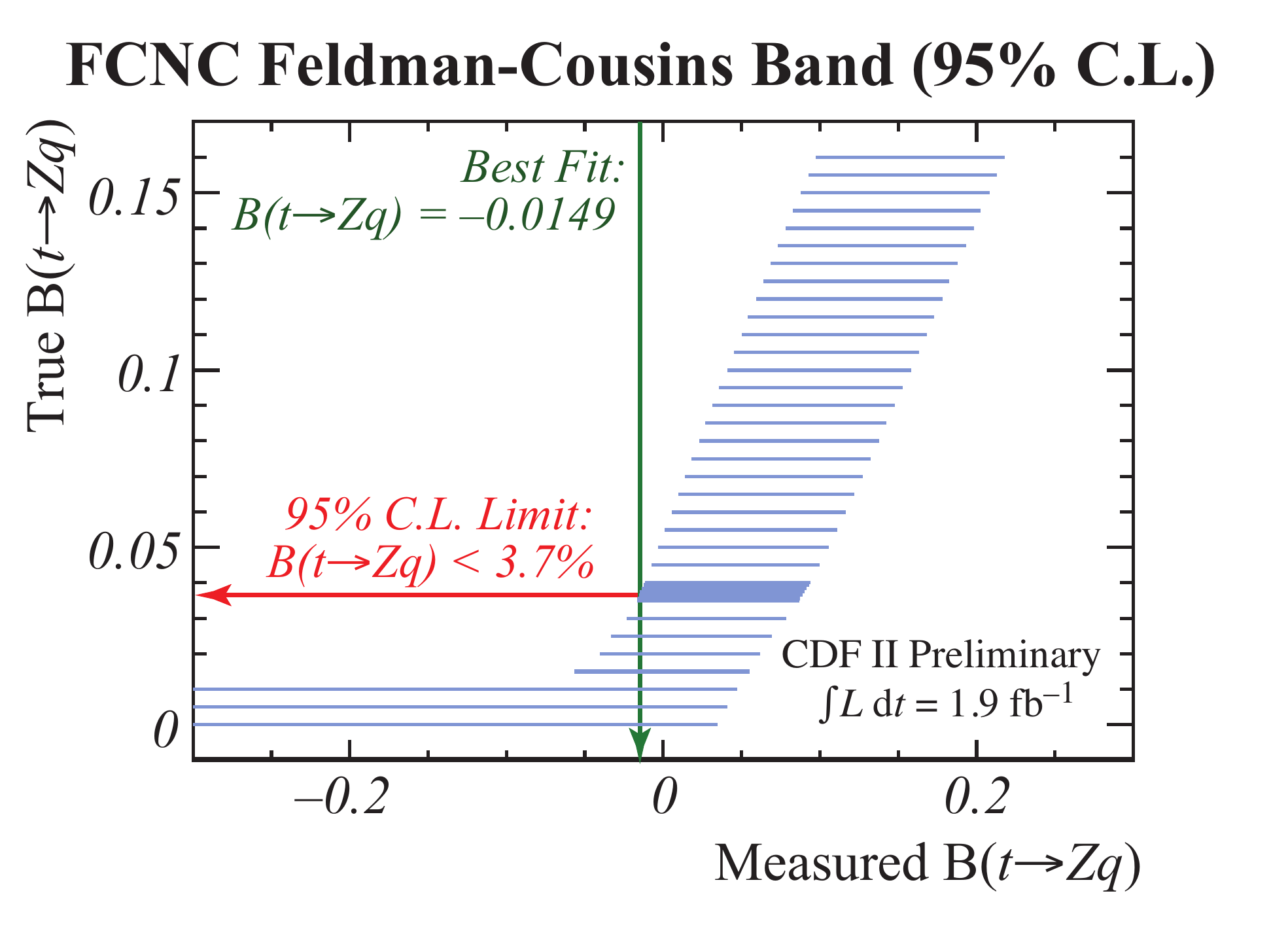}
\caption{\textbf{LEFT:}  $\chi^2$ distribution for two signal regions, with $\geq 1$ $b$-tag 
and no $b$-tags, and the control region.
\textbf{RIGHT:} Feldman-Cousins band at 95\% C.L. The measured value 
yields an upper limit of ${\cal B} (t \to Zq) < 3.7\%$ at 95\% C.L. 
}
\label{fig:FCNC}
 \end{figure}

This quantifies the consistency of each event with originating from a FCNC decay. 
Templates of this variable are generated for the main background, and for the FCNC signal.
Shape systematic uncertainties are included in the templates. The $\chi^2$ template fit 
is performed simultaneously in two signal regions and the control region. 
Results of the fit are consistent with the $\chi^2$ distribution for background alone. 
A top quark mass of 175  GeV/c$^2$ yields 
an upper limit of ${\cal B} (t \to Zq) < 3.7\%$ at 95\% C.L. based on the Feldman-Cousins
prescription~(see Fig.~\ref{fig:FCNC}).
%

\section{SEARCH FOR INVISIBLE TOP DECAYS}

The CDF Collaboration has also performed a generic search for alternative top decays by
measuring  the yield of the lepton $+$ jets events with two $b$-tags and quantifying 
the deficit from the value expected for the theoretical $\ttbar$ production 
cross section.  
By evaluating the relative acceptance ${\cal R}_{WX/WW}
 = {\cal A}(  \ttbar \to Wb XY ) /{\cal A}(  \ttbar \to Wb Wb )$,
 where $XY$ comprise non-standard decay,
and constructing Feldman-Cousins bands that relate the true
branching fraction for the considered decay to the number of 
observed lepton $+$ jets events,
CDF sets 95\% C.L. limits on the branching ratios. 
Results are listed in Table~\ref{tab:invisible} for several masses of the top quark.
Invisible top decays correspond to decay products that escape detection 
and do not contribute to the lepton $+$ jets final state.

\begin{table}[ht]
\begin{center}
\caption{95\% C.L. upper limits for branching ratios of top decays as a function of mass. }
\begin{tabular}{lcccc}
\hline \textbf{Decay} & \textbf{${\cal R}_{WX/WW}(\%)$} & \textbf{Limit  } &
\textbf{Limit  } &  \textbf{Limit } \\
  &  & (175 GeV) & (172.5 GeV) & (170 GeV) \\ \hline 
  $t \to Zc $ & 32 & 0.13 & 0.15 & 0.18 \\
   $t \to gc $ & 27 & 0.12 & 0.14 & 0.17 \\
    $t \to \gamma c $ & 18 & 0.11 & 0.12 & 0.15 \\
     $t \to $ invisible & 0 & 0.09 & 0.10 & 0.12 \\
\hline
\end{tabular}
\label{tab:invisible}
\end{center}
\end{table}
\section{CONCLUSIONS}

We presented most recent CDF and \Dzero precision measurements of the properties of the top quark. The results are in agreement with predictions of the SM.
Both experiments have set stringent limits on branching ratios for several decays beyond the SM.

\end{document}